\documentclass[useAMS,usenatbib]{mn2e}
\usepackage{times}
\usepackage[fleqn]{amsmath}
\usepackage{amssymb}
\usepackage{graphicx}
\usepackage{subfigure}
\usepackage{float}
\usepackage{relsize}
\usepackage{color}
\usepackage{gensymb}
\usepackage[mathscr]{euscript}

\title[Improved dust and gas mixtures in SPH]{Two-fluid dust and gas mixtures in smoothed particle hydrodynamics II: an improved semi-implicit approach}\author[Pablo Lor\'en-Aguilar and Matthew R. Bate]{Pablo Lor\'en-Aguilar$^{1}$\thanks{E-mail: pablo@astro.ex.ac.uk, mbate@astro.ex.ac.uk} and Matthew R. Bate$^{1}$ \\ $^{1}$ School of Physics and Astronomy, University of Exeter, Stocker Road, Exeter EX4 4QL, United Kingdom}

\date{Accepted 2015 September 28. Received 2015 September 8}

\begin{document}

\pagerange{\pageref{firstpage}--\pageref{lastpage}} \pubyear{???} \maketitle
\label{firstpage}

\begin{abstract} 
We present an improved version of the \cite{LB14} method to integrate the two-fluid dust/gas equations that correctly captures the limiting velocity of small grains in the presence
of net differences (excluding the drag force) between the accelerations of the dust and the gas. A series of accelerated \textsc{dustybox} tests and a simulation of dust-settling in a protoplanetary disc are performed comparing the performance of the new and old methods. The modified method can accurately capture the correct limiting velocity while preserving all the conservation properties of the original method.
\end{abstract}

\begin{keywords}
hydrodynamics -- methods: numerical -- planets and satellites: formation -- protoplanetary discs - dust, extinction.
\end{keywords}

\section{Introduction}

\begin{figure*} \centering
\includegraphics[width=80mm]{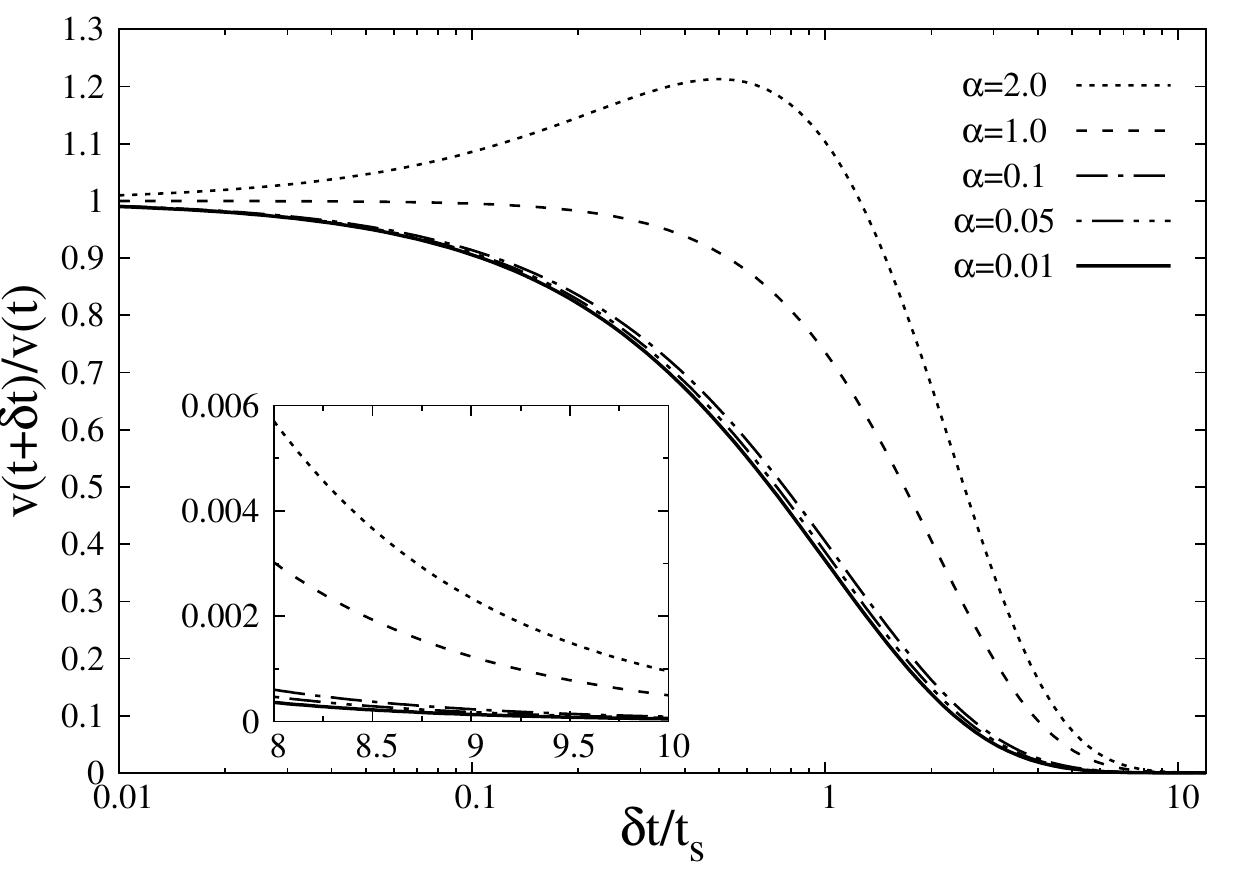} 
\includegraphics[width=80mm]{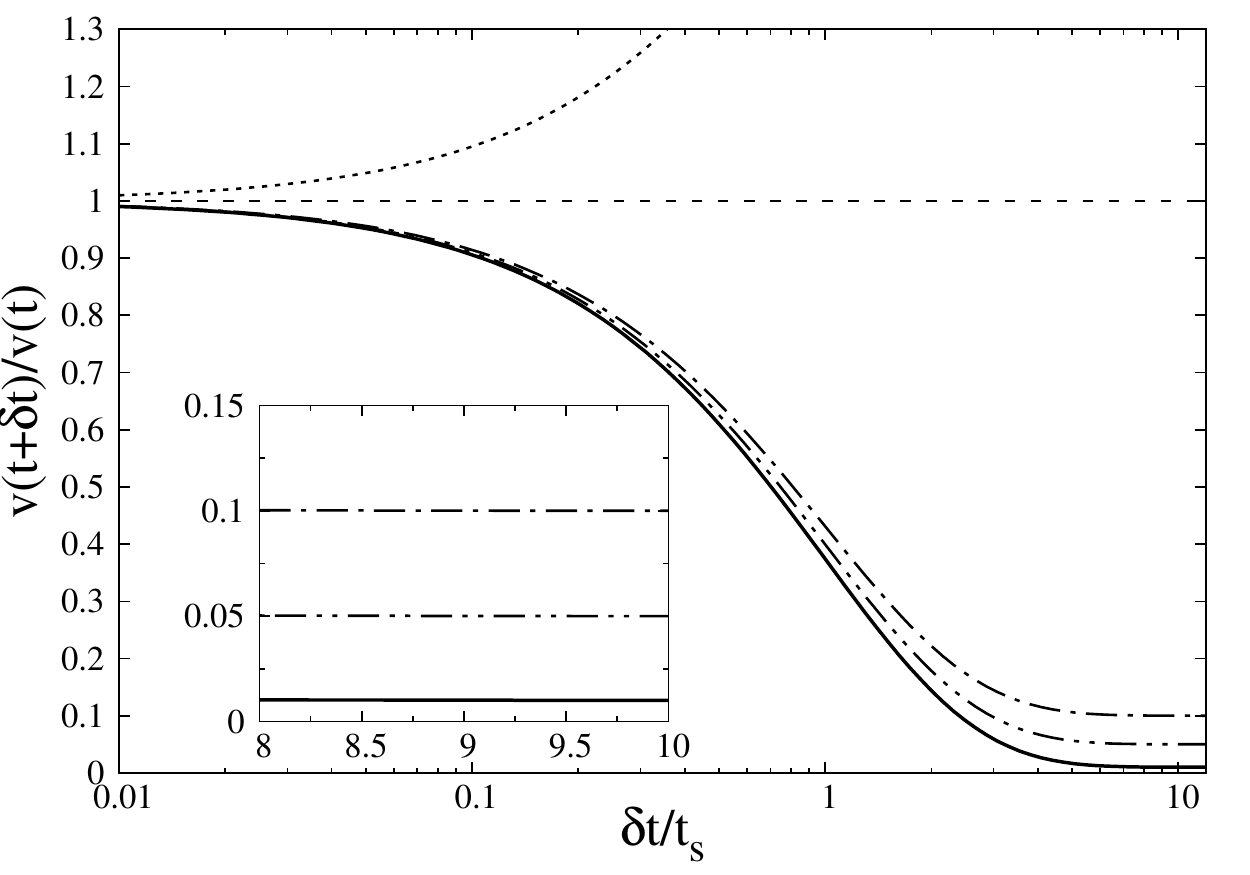} 
\caption{The evolution of the fractional change of the dust velocity after an integration time-step $\delta t$, for different values of the parameter $\alpha \equiv a_{\rm D} t_{\rm s}/v(t)$. The old method (left panel) produces excess drag in all cases, as can be seen when compared with the exact solution of the problem (right panel). For $\alpha \ll 1$, the solution from the old method is relatively close to the analytical solution. But for $\alpha \gtrsim 1$, one needs to enforce $\delta t/t_{\rm s} \lesssim 0.1$ in order to obtain close agreement with the analytical solution.  This is contrary to the original purpose of the method, which was to allow time-steps $\delta t \gg t_{\rm s}$ to be taken.}
\label{Fig0} \end{figure*}
A correct description of the evolution of dust and gas mixtures is essential to model many interesting astrophysical phenomena. One difficulty in modelling such mixtures numerically is the time-integration of the drag force between the gas and small dust grains \citep[see, e.g.][and references therein]{LB14}. The dust stopping time, $t_{\rm s}$, gives a measure of the time needed for the relative velocity of the dust with respect to the gas to reduce by a significant fraction. If the dust grain is very small, this time-scale may become exceedingly small in comparison with the gas evolutionary time-scale. As a consequence, a very large number of explicit time-steps need to be computed (or iterations in the case of implicit integration systems), making such simulations prohibitively expensive. 

To tackle this problem, a semi-implicit time integration method was proposed by \cite{LB14} in the framework of the smoothed particle hydrodynamics (SPH) \citep{Lucy1977,GM77} two-fluid scheme. Drag was implemented using an approximate solution of Euler's equations for the relative velocity of dust and gas
\begin{equation}
\textbf{v}_{\rm DG}(t+\delta t) = \textbf{v}_{\rm DG}(t)e^{-\delta t/t_{\rm s}}, \label{vDG_original}
\end{equation}
where $\textbf{v}_{\rm D}$ and $\textbf{v}_{\rm G}$ correspond to dust and gas velocities, respectively, $\textbf{v}_{\rm DG} \equiv \textbf{v}_{\rm D} - \textbf{v}_{\rm G}$, and $\delta t$ corresponds to the integration time-step. The scheme was implemented using an operator splitting technique. First, intermediate velocities were predicted for dust and gas components, excluding drag forces
\begin{align}
\tilde{\textbf{v}}_{\rm D}(t+\delta t) &= \textbf{v}_{\rm D}(t) + \textbf{a}_{\rm D}(t)\delta t, \label{vD_oldstep1}\\
\tilde{\textbf{v}}_{\rm G}(t+\delta t) &= \textbf{v}_{\rm G}(t) + \textbf{a}_{\rm G}(t)\delta t, \label{vG_oldstep1}
\end{align}
where $\textbf{a}_{\rm D}$ and $\textbf{a}_{\rm G}$ are the accelerations of the dust and gas, respectively, excluding drag forces. Subsequently, drag forces were applied using a time-discretized version of equation \ref{vDG_original}
\begin{align}
\textbf{v}_{\rm D}(t+\delta t) &= \tilde{\textbf{v}}_{\rm D}(t+\delta t) - \xi\tilde{\textbf{v}}_{\rm DG}(t+\delta t),  \label{vD_oldstep2}\\
\textbf{v}_{\rm G}(t+\delta t) &= \tilde{\textbf{v}}_{\rm G}(t+\delta t) + \epsilon\xi\tilde{\textbf{v}}_{\rm DG}(t+\delta t), \label{vG_oldstep2}
\end{align}
where $\epsilon \equiv \rho_{\rm D}/\rho_{\rm G}$ is the dust-to-gas ratio, and
\begin{align}
\xi &\equiv \frac{1-e^{-\delta t/t_{\rm s}}}{1+\epsilon}. \label{xi}
\end{align}
The method performed well in a variety of test cases. However, \cite{BSC15} recently pointed out a severe limitation of the method, namely that it does not produce the correct relative velocity between dust and gas for small grains in the presence of a net difference in the (non-drag) accelerations of the dust and the gas.  This occurs because the method is based on equation
\ref{vDG_original}, which is the solution of Euler's equations in the presence of drag without any additional
acceleration terms. Then, in the limit $\delta t/t_{\rm s} \rightarrow \infty$, $\xi \rightarrow 1/(1+\epsilon)$ and the application of equations \ref{vD_oldstep2} and \ref{vG_oldstep2} leads to $\textbf{v}_{\rm DG} \rightarrow 0$.  However, consider, for example, dust falling in a hydrostatic atmosphere.  In this case, both the dust and gas experience a gravitational acceleration, but for the gas this is balanced by the pressure gradient so that the net acceleration of the gas is zero, and $\textbf{a}_{\rm D}-\textbf{a}_{\rm G} = \nabla P_{\rm G}/\rho_{\rm G}$, where $P_{\rm G}$ is the gas pressure, and $\rho_{\rm G}$ the gas density.  In this case, the correct limiting velocity of the dust is $\textbf{v}_{\rm DG} \rightarrow t_{\rm s}\nabla P_{\rm G}/\rho_{\rm G}$, not zero.

In order to quantify the impact of the excessive drag produced by the use of equation \ref{vDG_original}, one can explore a  simple experiment. Consider a one-dimensional dust and gas mixture with initial velocities $\text{v}_{\rm D}=1$, $\text{v}_{\rm G}=0$, a constant acceleration $a_{\rm D}$ (affecting only the dust component) and an arbitrary dust stopping time $t_{\rm s}$. For simplicity, we take the dust-to-gas ratio $\epsilon \ll 1$. Using equations \ref{vD_oldstep1} to \ref{vG_oldstep2}, the velocity of the dust component can be evolved by a time-step $\delta t$ as
\begin{align}
\text{v}_{\rm D}(t+\delta t) &= \text{v}_{\rm D}(t) + \text{a}_{\rm D}\delta t - (1-e^{-\delta t/t_{\rm s}})\left(\text{v}_{\rm D}(t) +\text{a}_{\rm D}\delta t  \right) \nonumber \\
&=\left(\text{v}_{\rm D}(t) + \text{a}_{\rm D}\delta t\right) e^{-\delta t/t_{\rm s}}.
\end{align}
Thus, one can estimate the fractional change in the dust velocity after an integration time-step as
\begin{equation}
\frac{\text{v}_{\rm D}(t+\delta t)}{\text{v}_{\rm D}(t)} = \left(1 + \alpha\delta t\right) e^{-\delta t/t_{\rm s}}, 
\label{vD_fraction}\end{equation}
where $\alpha \equiv \text{a}_{\rm D}t_{\rm s}/\text{v}_{\rm D}(t)$. In Figure \ref{Fig0}, the fractional change in velocity as a function of $\delta t/t_{\rm s}$ is shown for various values of parameter $\alpha$, both for equation \ref{vD_fraction} (left panel), and for the correct analytical solution (right panel, see Section 2 for a derivation). In the left panel, the velocity at $t+\delta t$ always suffers an excess of drag with respect to the analytical solution. For $\alpha \ll 1$, equation \ref{vD_fraction} produces only a very small fractional error. The problem occurs for values $\alpha \gtrsim 1$, when the predicted value for the velocity completely diverges from the analytical solution if $\delta t  >  0.1t_{\rm s}$. Physically this can only occur if the change of the velocity produced by the acceleration is large enough, i.e. $at_{\rm s} \gtrsim 0.1 \text{v}_{\rm D}$. Such a circumstance should not normally occur if the non-drag accelerations are shared by both dust and gas components, since the gas time-step condition should automatically restrict $\delta t$. However, such a restriction will not occur, for example, if an acceleration is only felt by the dust component. Then, depending on the specific values of the dust acceleration, stopping time and velocity, the result may be completely wrong.

In the case of dust falling in a hydrostatic atmosphere, because the total acceleration of the gas component is close to zero, the time-step $\delta t$ may have little to do with the 
gravitational acceleration or the velocity of the settling dust grains. The error for weakly coupled dust grains will be very small, since $\delta t/t_{\rm s} \ll 1$. Similarly, the error for strongly coupled grains will be very small, since the relative velocity will be very small, leading to a small absolute error despite the very big fractional error. However, intermediately coupled grains may simultaneously generate sizeable fractional and absolute errors. Hence, the only way to recover the correct evolution, independently of $t_{\rm s}$, is to force $\delta t/t_{\rm s} \lesssim 0.1$ by reducing the integration time-step $\delta t$. This restriction is clearly in conflict with the purpose of the originally designed algorithm, i.e. avoiding the time-stepping restriction of the dust force. Hence, a modification of the method to recover the proper limits when $\delta t/t_{\rm s} \rightarrow \infty$ is mandatory. 

In this paper, we present an improved version of the method that produces the appropriate limiting velocities of the small grains in the presence of accelerations. In Section 2, the modified numerical method is presented, in Section 3 we present the results of numerical tests and, finally, in Section 4 we draw our conclusions. 

\section{Numerical method}
\label{numerical}

Euler's equations can be expressed as a function of the relative and barocentric velocities as \citep[e.g][]{YG05,LP14}
%
%
\begin{align}
\mathscr{D}_{\rm t}\textbf{v}_{\rm DG}  &= \textbf{a}_{\rm DG} - \frac{\textbf{v}_{\rm DG}}{t_{\rm s}}  - \left(\textbf{v}_{\rm DG}\cdot\nabla\right)\textbf{v} - \textbf{G}(\textbf{v}_{\rm DG}^2),  \label{Eu_vDG} \\
\mathscr{D}_{\rm t}\textbf{v} &= \textbf{a}  - \textbf{F}(\textbf{v}_{\rm DG}^2) \label{Eu_v}, \end{align}
where we emphasise that $\textbf{a}_{\rm G}$ includes accelerations due to gas pressure gradients, and we define the total density $\rho = \rho_{\rm D} + \rho_{\rm G}$, which includes the dust density $\rho_{\rm D}$, $\textbf{v} \equiv \left(\rho_{\rm D}\textbf{v}_{\rm D} + \rho_{\rm G}\textbf{v}_{\rm G}\right)/\rho$, $\textbf{v}_{\rm DG} \equiv \textbf{v}_{\rm D}-\textbf{v}_{\rm G}$, $\textbf{a} \equiv \left(\rho_{\rm D}\textbf{a}_{\rm D} + \rho_{\rm G}\textbf{a}_{\rm G}\right)/\rho$,  $\textbf{a}_{\rm DG} \equiv \textbf{a}_{\rm D}-\textbf{a}_{\rm G}$, and
\begin{align}
\textbf{F}(\textbf{v}_{\rm DG}^2) &\equiv \frac{1}{\rho}\nabla\cdot\left(\frac{\rho_{\rm D}\rho_{\rm G}}{\rho}\textbf{v}_{\rm DG}^2\right),\\
\textbf{G}(\textbf{v}_{\rm DG}^2) &\equiv \frac{\rho_{\rm G}}{\rho}\textbf{v}_{\rm DG}\cdot\nabla\left(\frac{\rho_{\rm G}}{\rho}\textbf{v}_{\rm DG}\right) - \frac{\rho_{\rm D}}{\rho}\textbf{v}_{\rm DG}\cdot\nabla\left(\frac{\rho_{\rm D}}{\rho}\textbf{v}_{\rm DG}\right).
\end{align}
The stopping time can be expressed as $t_{\rm s} \equiv \rho_{\rm G}\hat{m}_{\rm D}/(K_{\rm s}\rho)$, where $\hat{m}_{\rm D}$ is the mass of a single dust grain, and $K_{\rm s}$ is its drag coefficient.
\begin{figure*} \centering
\includegraphics[width=80mm]{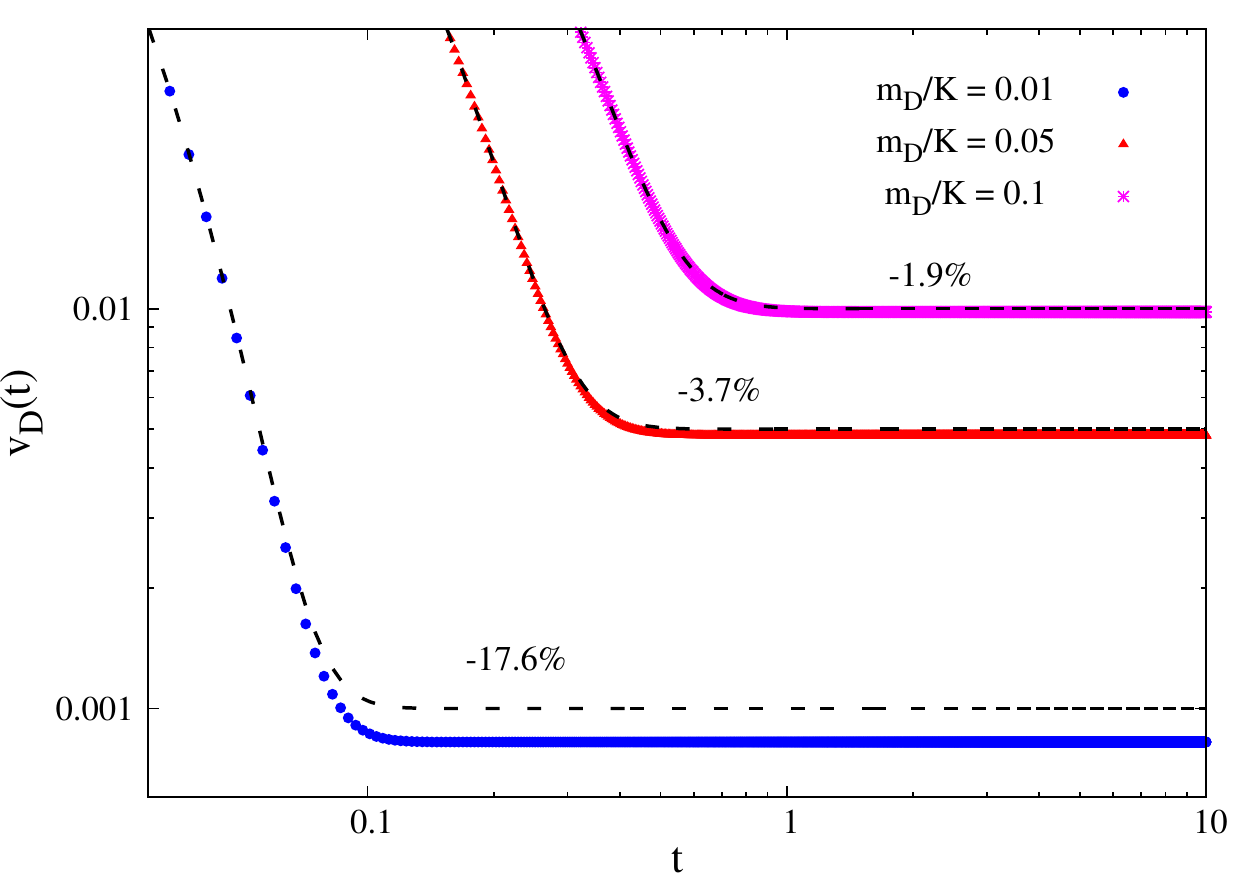}
\hspace{3mm}\includegraphics[width=78mm,height=56mm]{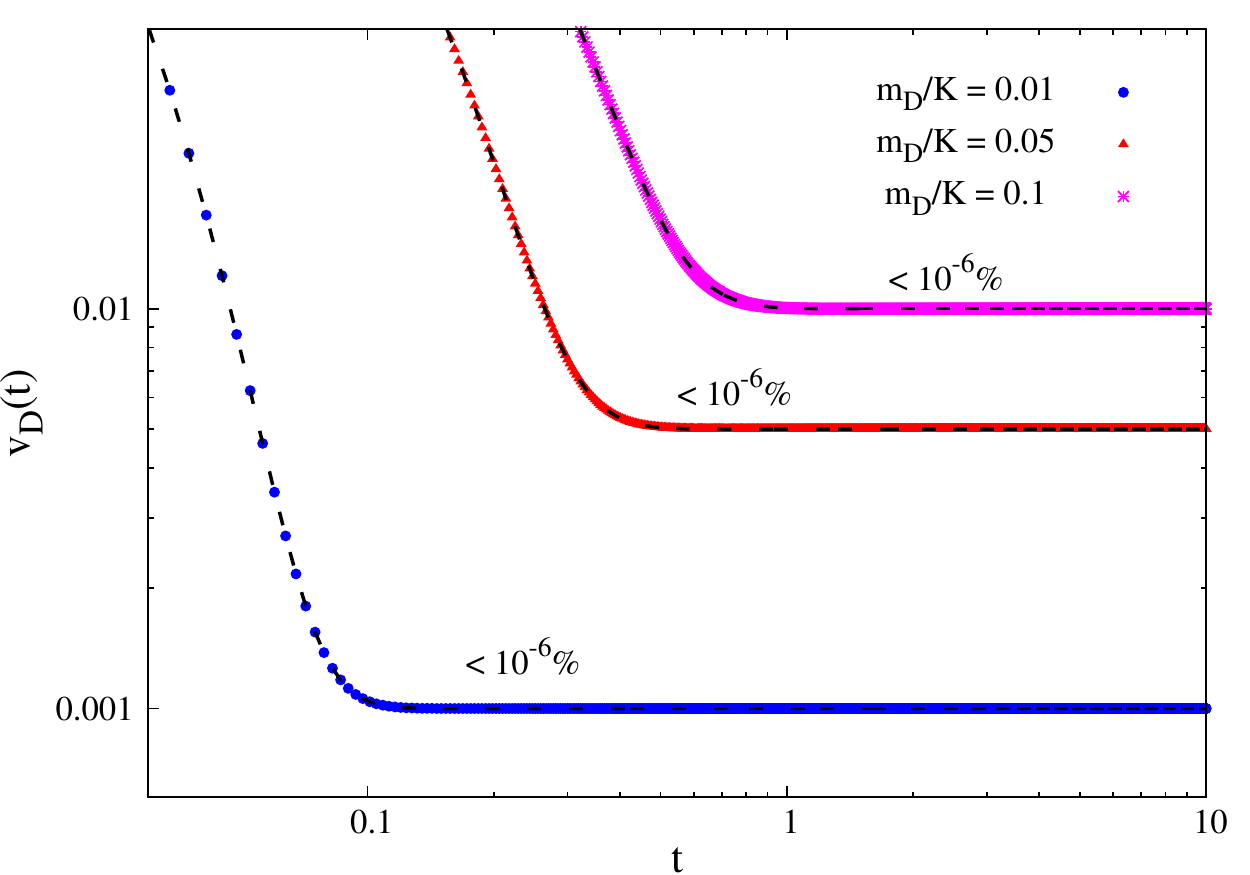}
\caption{The time evolution of the velocity of a dust particle experiencing a constant acceleration and gas drag, in the test particle limit ($\rho_{\rm D}/\rho_{\rm G} \ll 1$), for different values of the drag strength coefficient $\hat{m}_{\rm D}/K_{\rm s}$ in one-dimensional calculations. The new method (right panel) predicts the terminal velocity with a very high degree of accuracy, whereas the old method (left panel) produces the expected excess of drag. The label above each solution indicates the relative error of the numerical solution compared to the analytical solution.} \label{Fig1} \end{figure*}
If (i) densities and accelerations can be considered approximately constant during the integration time-step, and (ii) the relative advection terms, \textbf{F} and \textbf{G}, can be neglected due to the smallness of $\textbf{v}_{\rm DG}$ (see \cite{YG05} and \cite{BSC15} for a discussion), the solution of equations \ref{Eu_vDG} and \ref{Eu_v} can be written as
\begin{align}
\textbf{v}_{\rm DG}(t+\delta t) &= \textbf{v}_{\rm DG}(t) e^{-\delta t/t_{\rm s}} + \textbf{a}_{\rm DG}\left(1-e^{-\delta t/t_{\rm s}}\right)t_{\rm s} \label{v_sol},\\
\textbf{v}(t+\delta t) &= \textbf{v}(t) +\textbf{a} \delta t. \label{vDG_sol} 
\end{align}
Using
\begin{align}
\textbf{v}_{\rm D} &= \textbf{v} + \frac{\rho_{\rm G}}{\rho}\textbf{v}_{\rm DG}, \label{vD}\\
\textbf{v}_{\rm G} &= \textbf{v} - \frac{\rho_{\rm D}}{\rho}\textbf{v}_{\rm DG}, \label{vG}
\end{align}
equations \ref{v_sol} and \ref{vDG_sol} can be implemented using again a two-step method. As in the original method we perform a standard explicit integration to apply non-drag forces (equations \ref{vD_oldstep1} and \ref{vG_oldstep1}), but subsequently we apply the drag forces using
\begin{align}
\textbf{v}_{\rm D}(t+\delta t) &= \tilde{\textbf{v}}_{\rm D}(t+\delta t) - \xi\tilde{\textbf{v}}_{\rm DG}(t+\delta t) +\Lambda\textbf{a}_{\rm DG}(t), \label{vD_step2}\\
\textbf{v}_{\rm G}(t+\delta t) &= \tilde{\textbf{v}}_{\rm G}(t+\delta t) + \epsilon\xi\tilde{\textbf{v}}_{\rm DG}(t+\delta t) -\epsilon\Lambda\textbf{a}_{\rm DG}(t), \label{vG_step2}
\end{align}
where 
\begin{align}
\Lambda &\equiv (\delta t + t_{\rm s})\xi - \frac{\delta t}{1+\epsilon}. \label{lambda}
\end{align}
To calculate the time-evolution of the internal energy of the gas $u_{\rm G}$, one can make use of energy conservation. The total change in kinetic energy per unit volume of the mixture will be given by
\begin{align}
\Delta\text{E}_{\rm K} & =   \frac{1}{2}\rho_{\rm D}\textbf{v}^2_{\rm D}(t+\delta t)+\frac{1}{2}\rho_{\rm G}\textbf{v}^2_{\rm G}(t+\delta t) \nonumber \\
& \mbox{\hspace{0.4cm}}     - \frac{1}{2}\rho_{\rm D}\textbf{v}^2_{\rm D}(t)-\frac{1}{2}\rho_{\rm G}\textbf{v}^2_{\rm G}(t) \nonumber \\
&=  \frac{1}{2}\rho_{\rm D}\left(\tilde{\textbf{v}}^2_{\rm D}(t+\delta t_{\rm G})-\textbf{v}^2_{\rm D}(t)\right)+\frac{1}{2}\rho_{\rm G}\left(\tilde{\textbf{v}}^2_{\rm G}(t+\delta t) \right. \nonumber \\
& \mbox{\hspace{0.4cm}}      \left.-\textbf{v}^2_{\rm G}(t)\right) -\rho_{\rm D}\left(\tilde{\textbf{v}}_{\rm DG}(t+\delta t) - \frac{1}{2}\left(1+\epsilon\right)\textbf{S}_{\rm DG}\right)\textbf{S}_{\rm DG} \nonumber \\
&=  \Delta \tilde{\text{E}}_{\rm K} - \rho_{\rm D}\left(\tilde{\textbf{v}}_{\rm DG} (t+\delta t)- \frac{1}{2}\left(1+\epsilon\right)\textbf{S}_{\rm DG}\right)\textbf{S}_{\rm DG}
\end{align}
where
\begin{align}
\textbf{S}_{\rm DG} \equiv \xi\tilde{\textbf{v}}_{\rm DG}(t+\delta t) - \Lambda\textbf{a}_{\rm DG}(t),
\end{align}
and $\Delta \tilde{\text{E}}_{\rm K}$ is the total change in kinetic energy per unit volume due to non-drag forces. So, assuming that the total change in thermal energy is given by the total lost kinetic energy
\begin{align}
u_{\rm G}(t+\delta t) &= \tilde{u}_{\rm G}(t+\delta t) \nonumber \\
&+ \frac{\rho_{\rm D}}{\rho_{\rm G}}\left(\tilde{\textbf{v}}_{\rm DG} (t+\delta t)- \frac{1}{2}\left(1+\epsilon\right)\textbf{S}_{\rm DG}\right)\textbf{S}_{\rm DG}. \label{uG_step2} 
\end{align}

\begin{figure*} \centering
\hspace{0.6mm}\includegraphics[width=80mm]{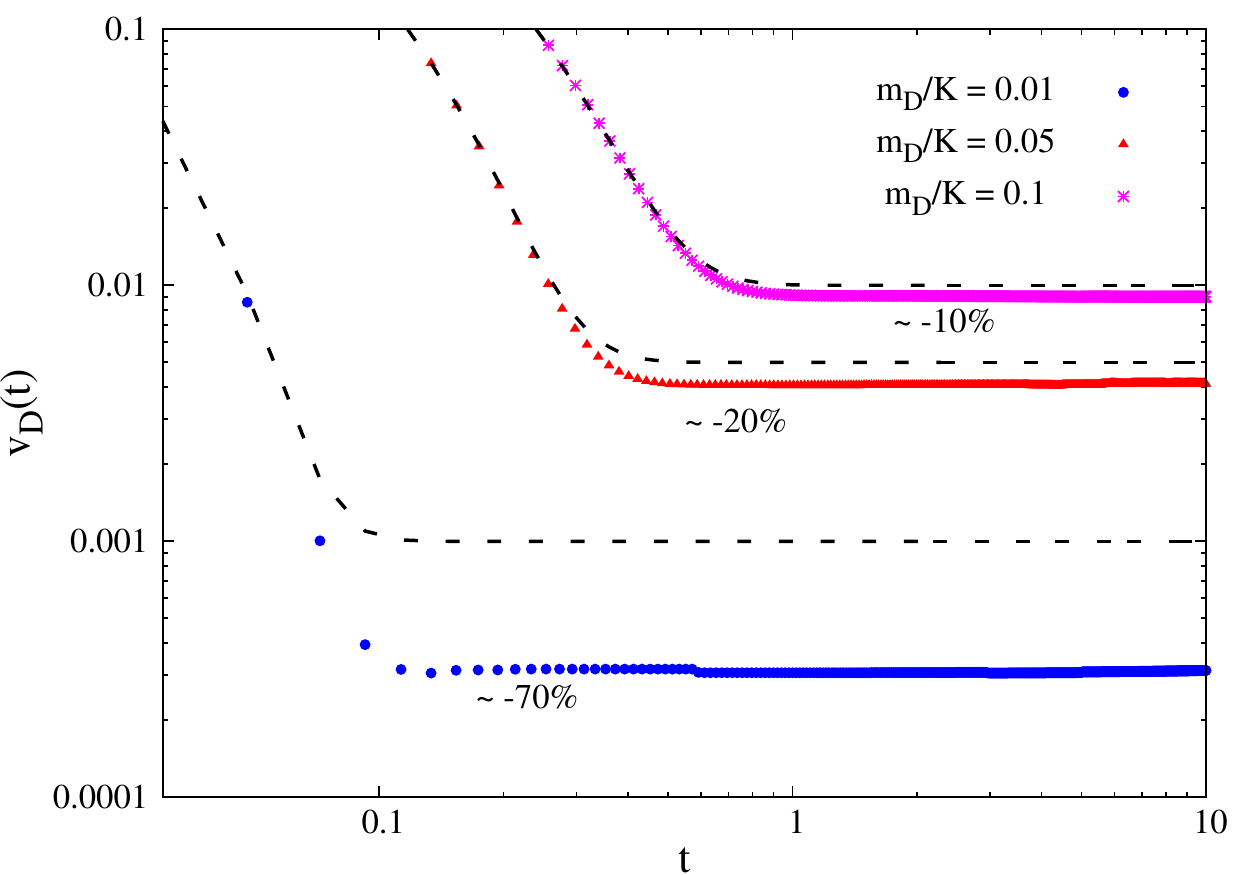}
\hspace{3mm}\includegraphics[width=80mm]{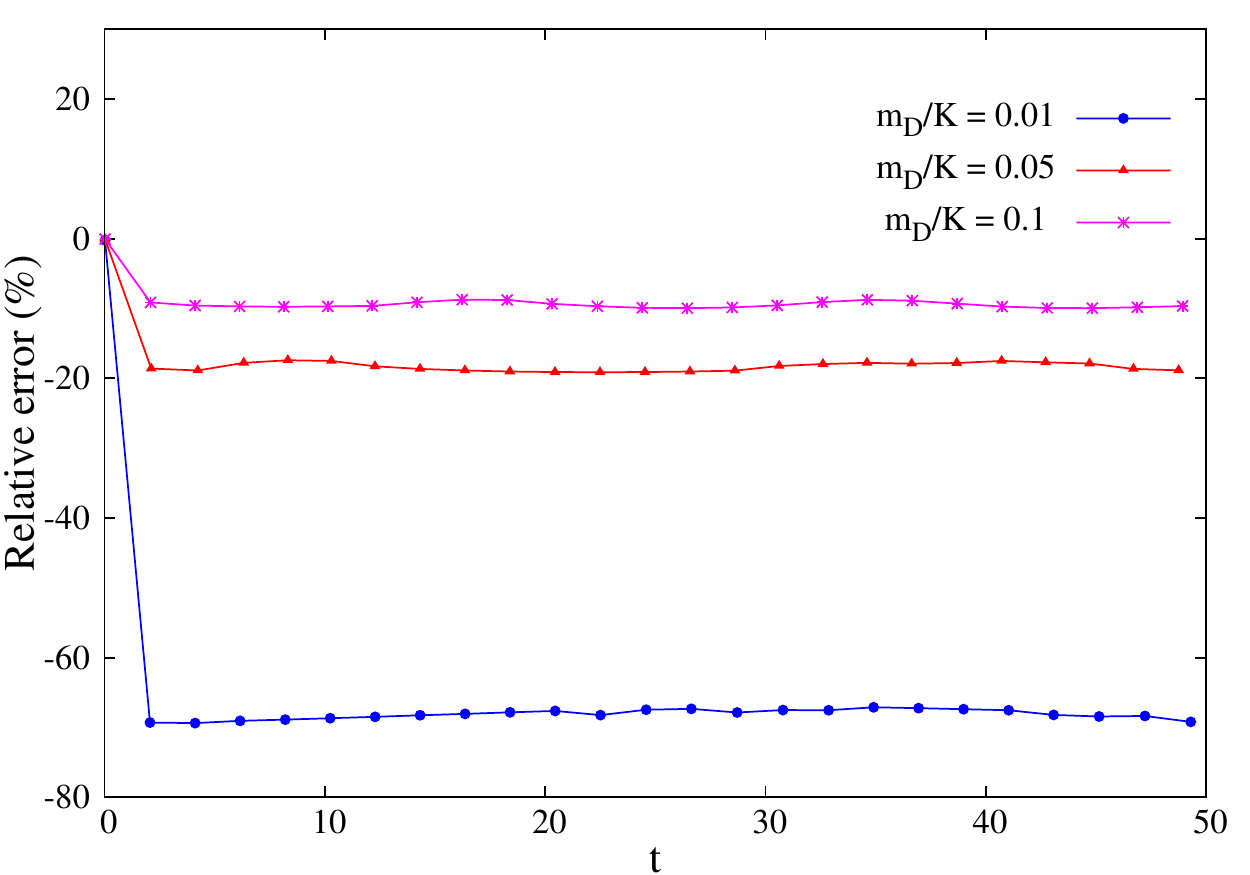}\\
\includegraphics[width=80mm]{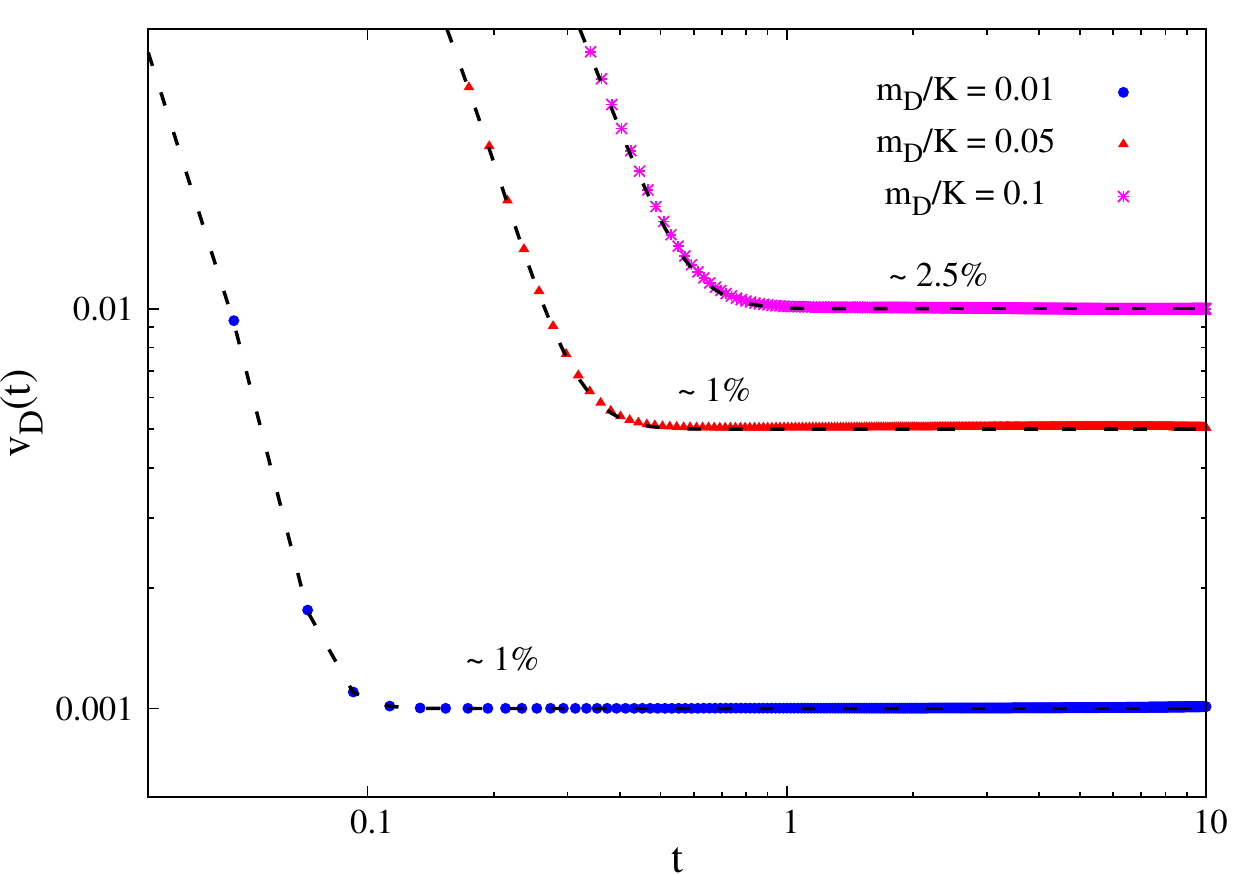}
\hspace{3mm}\includegraphics[width=80mm]{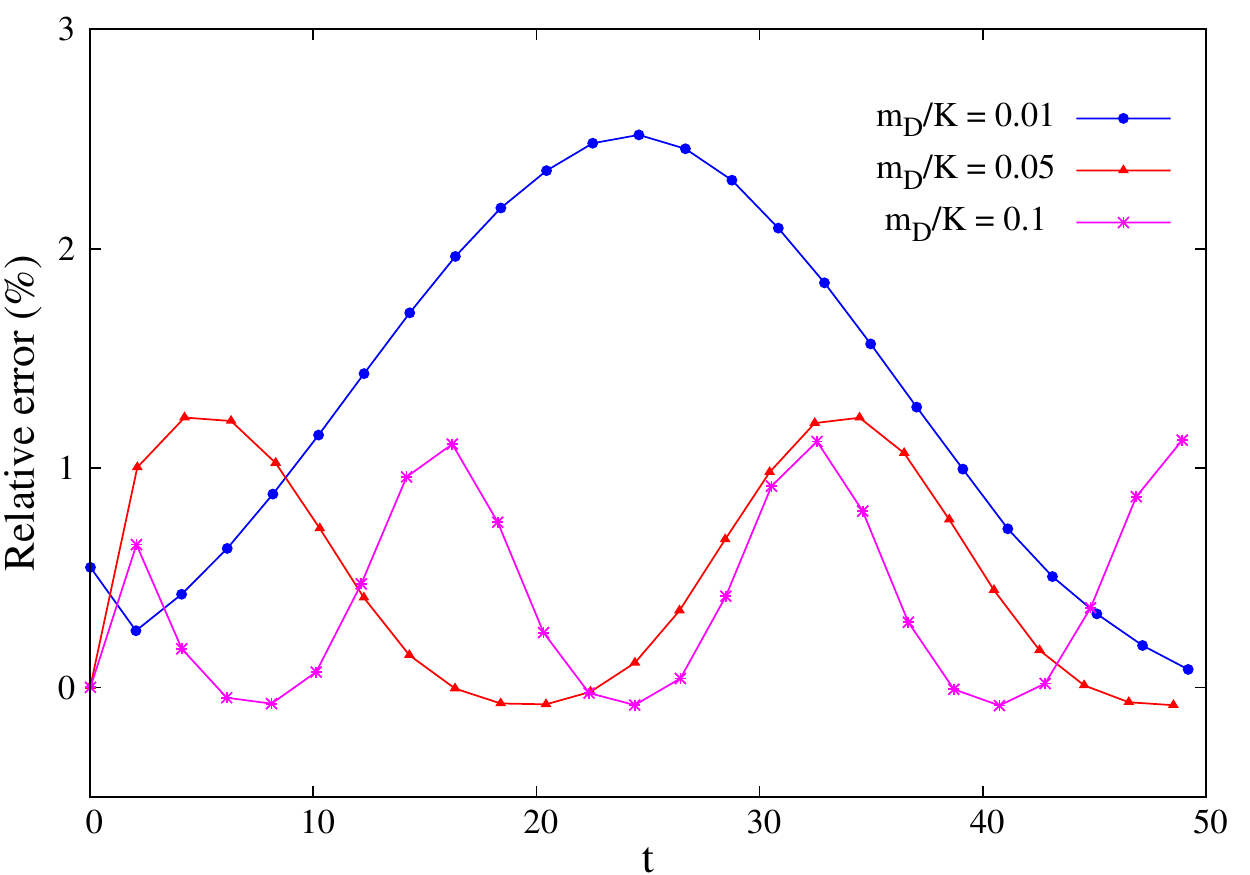}
\caption{The time evolution of the velocity of a dust particle experiencing a constant acceleration and gas drag, in the test particle limit ($\rho_{\rm D}/\rho_{\rm G} \ll 1$), for different values of the drag strength coefficient $\hat{m}_{\rm D}/K_{\rm s}$ in three-dimensional calculations. The old method (upper panels) produces excess drag, whereas the new method (lower panels) achieves the correct terminal velocity to a good degree of accuracy. The label close to each solution in the left panels indicates the relative error of the numerical solution compared to the analytical solution. As explained in the main text, fluctuations in the terminal velocity (lower right panel) arise as consequence of the underlying grid structure of the gaseous component, due to the finite resolution.} \label{Fig2}
\end{figure*}

We can check our equations produce the expected behaviour in the limits of small and large time-steps. If $\delta t/t_{\rm s} \ll 1$, $\xi \rightarrow \delta t /t_{\rm s}(1+\epsilon)$, $\Lambda \rightarrow \delta t^2/t_{\rm s}(1+\epsilon)$, so equations \ref{vD_step2} and \ref{vG_step2} become
\begin{align}
\textbf{v}_{\rm D}(t+\delta t) &= \textbf{v}_{\rm D}(t) - \frac{\textbf{v}_{\rm DG}(t)}{t_{\rm s}}\delta t +\textbf{a}_{\rm D}(t)\delta t, \label{vD_explicit}\\
\textbf{v}_{\rm G}(t+\delta t) &= \textbf{v}_{\rm G}(t) + \epsilon\frac{\textbf{v}_{\rm DG}(t)}{t_{\rm s}}\delta t +\textbf{a}_{\rm G}(t)\delta t, \label{vG_explicit}
\end{align}
recovering the low-drag explicit integration regime. On the other hand, if $\delta t/t_{\rm s} \gg 1$, $\xi \rightarrow 1/(1+\epsilon)$, $\Lambda \rightarrow t_{\rm s}/(1+\epsilon)$, and equations \ref{vD_step2} and \ref{vG_step2} become
\begin{align}
\textbf{v}_{\rm D}(t+\delta t,\textbf{r}_{\rm D}) &= \textbf{v}(t,\textbf{r}_{\rm D}) + \textbf{a}(t,\textbf{r}_{\rm D})\delta t + \frac{t_{\rm s}}{1+\epsilon}\textbf{a}_{\rm DG}(t,\textbf{r}_{\rm D}), \label{vD_impl}\\
\textbf{v}_{\rm G}(t+\delta t,\textbf{r}_{\rm G}) &= \textbf{v}(t,\textbf{r}_{\rm G}) + \textbf{a}(t,\textbf{r}_{\rm G})\delta t - \frac{\epsilon}{1+\epsilon} t_{\rm s}\textbf{a}_{\rm DG}(t,\textbf{r}_{\rm G}), \label{vG_impl}
\end{align}
recovering the appropriate strong drag limit, in which $\textbf{v}_{\rm DG} = t_{\rm s}\textbf{a}_{\rm DG}$. Equations \ref{vD_step2}, \ref{vG_step2} and \ref{uG_step2} can be implemented in the SPH method using the discretization procedure discussed in \cite{LB14}.  This gives
\begin{align} \textbf{v}^{i}_{\rm D}(t+\delta t,
\textbf{r}_{i})&= \tilde{\textbf{v}}^{i}_{\rm D}(t+\delta t,\textbf{r}_{i}) \nonumber \\ 
&-\frac{\nu}{N_{i}}\sum_{k}^{\rm Gas}\frac{m_{k}} {\rho_{k}}
\left(\textbf{S}_{ik}\cdot\hat{\textbf{r}}_{ik}\right)\hat{\textbf{r}}_{ik} W(|\textbf{r}_{ik}|,h_{k}),  \label{vD_SPH1} 
 \end{align}
\begin{align} 
\textbf{v}^{j}_{\rm G}(t+\delta t,\textbf{r}_{j}) &= \tilde{\textbf{v}}^{j}_{\rm G}(t+\delta t,\textbf{r}_{j}) \nonumber \\
&+\nu\sum_{k}^{\rm Dust}\frac{m_k}{N_{k}\rho_{j}}
(\textbf{S}_{kj}\cdot\hat{\textbf{r}}_{kj})\hat{\textbf{r}}_{kj}
W(|\textbf{r}_{kj}|,h_{j}),  \label{vG_SPH1} 
 \end{align}
\begin{align} 
u^{j}_{\rm G} (t+\delta t,\textbf{r}_{j}) &=  \tilde{u}^{j}_{\rm G} (t +\delta t,\textbf{r}_{j}) \nonumber \\
& + \sum_{k}^{\rm Dust} \frac{m_{k}}{N_{k}\rho_{k}}\left[ \begin{array}{c}~ \\ ~ \end{array} \hspace{-12pt} \left(\textbf{S}_{kj}\cdot\hat{\textbf{r}}_{kj}\right)\left(\textbf{v}_{kj}\cdot\hat{\textbf{r}}_{kj}\right) W(|\textbf{r}_{kj}|,h_{j}) \right. \nonumber \\
&- \left.\frac{1}{2}\left(1+\rho_{k}/\rho_{j}\right)\left(\textbf{S}_{kj}\cdot\hat{\textbf{r}}_{kj}\right)^2 W(|\textbf{r}_{kj}|,h_{j})\right] \label{uG_SPH1} 
\end{align}
where $\nu$ is the number of spatial dimensions, $\textbf{r}_{i}$ is the position of the $i$th particle, $\textbf{r}_{ik} \equiv \textbf{r}_{i} - \textbf{r}_{k}$, $m_{k}$ and $h_{k}$ are the mass and smoothing lengths of the $k$th particle, respectively, and $W$ is the interpolating function, known as the SPH kernel \citep[see for example][]{MN}.  We have also included a normalisation factor for the dust
\begin{equation} 
N_{i}  \equiv \sum_{k}^{\rm Gas}\frac{m_{k}}{\rho_{k}}W(|\textbf{r}_{ik}|,h_{k}). \label{norm}
\end{equation}
Equation \ref{uG_SPH1} assumes that all the kinetic energy dissipated by drag is transformed into thermal energy of the gas (see \cite{LB14} for a detailed explanation of the procedure).

\section{Results and discussion}
\label{results}

\begin{figure*} \centering
\includegraphics[height=62mm]{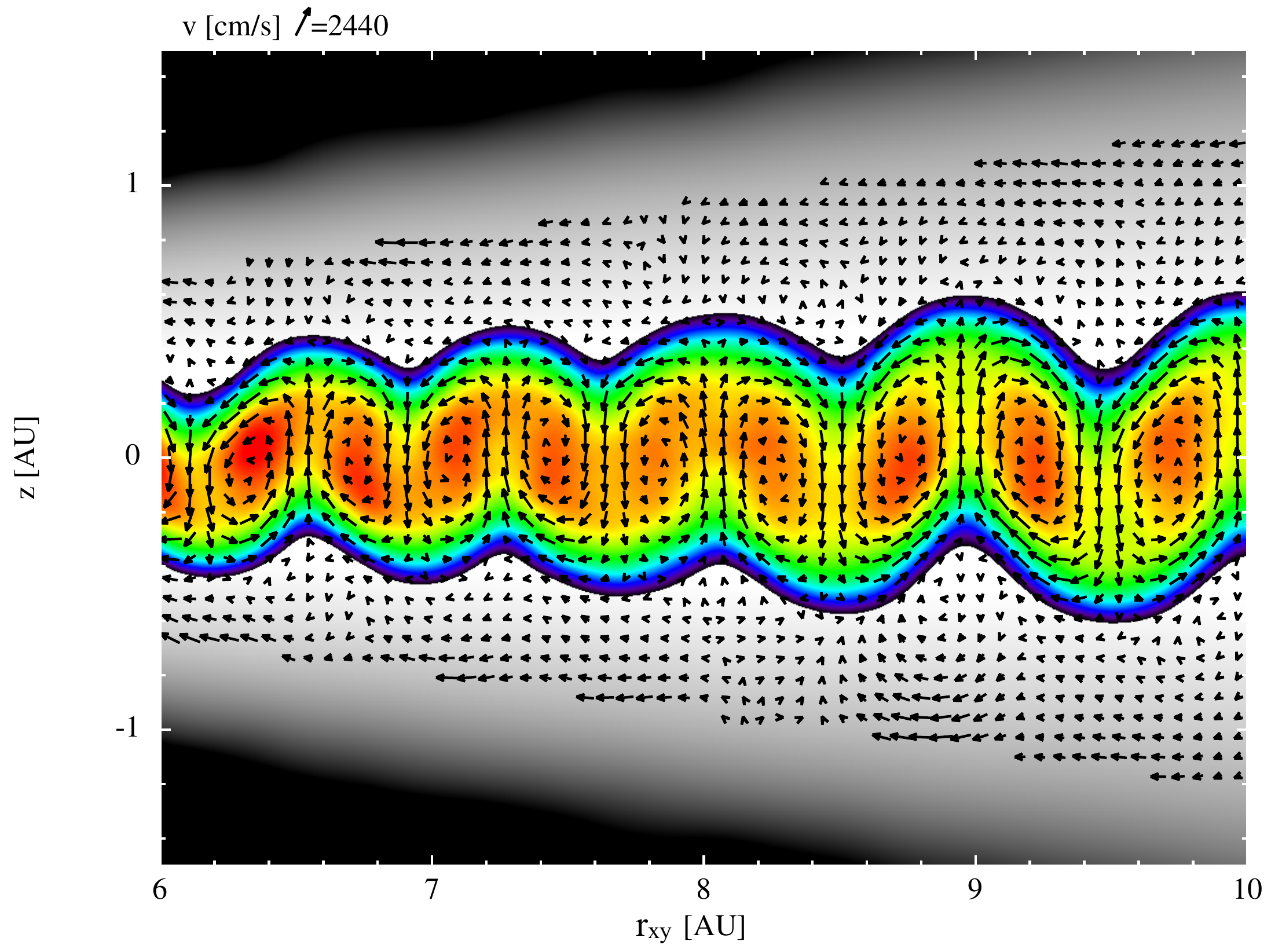}
\hspace{3mm} \includegraphics[height=62mm]{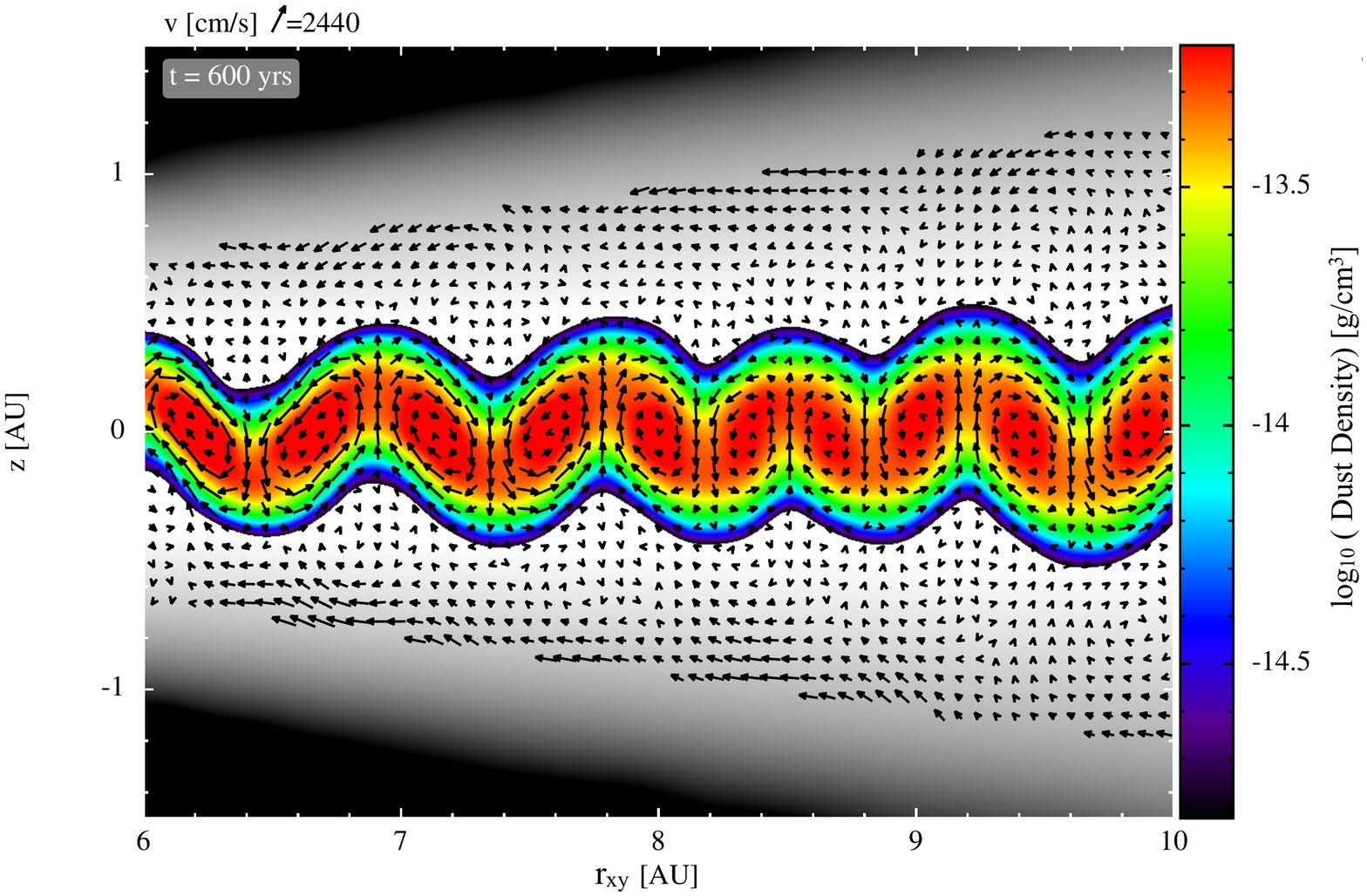}
\caption{Azimuthally-averaged density rendering of the dust (colour scale) and gas (greyscale, ranging over $\log_{10}(\rho_{\rm G}) = [-17, -12]$ in g~cm$^{-3}$) components of a protoplanetary disc. Vectors represent the velocity field of the fluid. The dust component of the disc is comprised of 1~mm dust grains with an initial dust-to-gas ratio of $\epsilon = 0.01$. After the dust grains settle, a compositional baroclinic instability develops creating toroidal vortices. The left panel is taken from \protect\cite{LB15}, who used the original semi-implicit integration method of \protect\citep{LB14} which produces excess drag and a relatively thick dust layer. Using the improved integration method (right panel) the instability is qualitatively the same, but the dust layer is somewhat thinner.} \label{disc} \label{Fig3} \end{figure*}

To test the accuracy of the improved algorithm, a series of \textsc{dustybox} experiments were performed. A set of 128 particles per phase in one dimension and $20^3$ particles per phase in three dimensions with homogeneous densities $\rho_{\rm G}$ and $\rho_{\rm D}$ were placed in a periodic box with an initial velocity $\textbf{v}_{\rm D}=(1,0,0)$ and $\textbf{v}_{\rm G}=(0,0,0)$. A constant acceleration $\textbf{a}_{\rm D}=(0.1,0,0)$ was applied exclusively to the dust particles.  To construct the initial 3D model, particles were evenly distributed in a cubic lattice with $-2 \le x,y,z \le 2$. The dust lattice was shifted, with respect to the gas, by half of the gas particles separation in each direction. The mass of each SPH particle was equal to $m_{\rm G} = V\rho_{\rm G}/N$ and $m_{\rm D} = \epsilon m_{\rm G}$, where $V$ is the volume of the computational domain, and $N$ is the number of particles in each phase.

The time evolution of the velocity of an arbitrary dust particle should be given by
\begin{equation}
\textbf{v}_{\rm D}(t) = \textbf{v}_{\rm D}(0)(1-\xi(t)) + \left(\frac{\epsilon}{1+\epsilon} t + \xi(t) t_{\rm s}\right)\textbf{a}_{\rm D},  \label{vD_sol}
\end{equation}
where
\begin{align}
\xi(t) &\equiv \frac{1-e^{- t/t_{\rm s}}}{1+\epsilon}. \label{xi}
\end{align}
In the test particle limit (i.e. $\epsilon \ll 1$), dust particles should reach a constant limiting velocity given by $\textbf{v}_{\rm D} (t \rightarrow \infty) = t_{\rm s}\textbf{a}_{\rm D}$. In the right panel of Fig.\ \ref{Fig1}, the time evolution of the velocity for an arbitrary dust grain is shown for various stopping times in a one-dimensional case. In the left panel of Fig.\ \ref{Fig1}, the results obtained with the original \cite{LB14} method are shown. In each of these calculations the densities have been taken as constants (rather than using SPH summations). As expected, using the old method, the terminal velocity is incorrectly predicted due to the expected drag excess caused by $\xi(\delta t/t_{\rm s} \rightarrow \infty)\rightarrow 1/(1+\epsilon)$. Using the new method, the limiting velocity is correctly predicted within high accuracy. 

In Fig.\ \ref{Fig2}, the results obtained using the original and improved methods are shown for a three-dimensional case. This time, densities are self-consistently calculated using SPH summations as dust and gas particles evolve in time. In the lower panels, the obtained result using the improved method is shown. Again, the correct terminal velocities are obtained with a good degree of accuracy.
As shown in the right panels, small fluctuations (of order 1 percent) are observed in the limiting velocity of the dust particles. \cite{BSC15} speculated that similar fluctuations in \cite{LB14} may be due to the original method's inability to model the correct limiting velocity, but this is not the case.  The fluctuations occur as a result of the motion of the dust particles relative to the grid of gas particles, which are essentially motionless (as $\epsilon \ll 1$). Consequently, the acceleration of the dust particle undergoes small periodic changes as it travels through the box. The spacing between the gas particles is 0.2 code units. Consequently, dust particles travelling at limiting velocities of 0.01, 0.005 and 0.001 in code units should expect to find along its way gas particles at time intervals of the order of 20, 40 and 200 code units, respectively. This is approximately the periodicity of the velocity fluctuations seen in the right panels of Fig.\ \ref{Fig2}.  We note that \cite{BSC15} propose using a method that does not use pair-wise forces between dust and gas particles and because they only study dust/gas drag in the test particle limit, they do not include the back reaction of the dust drag on the gas.  By contrast, our method includes the back reaction and guarantees momentum conservation.

Finally, we applied the improved method to the simulation of a realistic astrophysical problem. \cite{LB15} reported results from simulations of dust settling in protoplanetary discs, and discovered a new type of instability whereby vertical gradients in the dust-to-gas ratio drive a baroclinic instability that produces toroidal gas vortices.  The instability manifests itself for intermediate size dust grains ($\sim 1$~mm in the reported calculations) that can undergo vertical settling, but are nevertheless quite well coupled to the gas.  Since the onset of the instability critically depends on the settling velocity of the grains and the associated gradients in the dust-to-gas ratio, any inaccuracy in the calculation of the limiting velocity of the grains could lead to differences in the evolution of the instability. Therefore, we have repeated the main calculation of \cite{LB15} (see their paper for further details of the set up and initial conditions) using our improved method to investigate the impact of the excess drag on the earlier results. In Fig.\ \ref{Fig3}, we compare the dust distribution and toroidal vortices from original calculation (left panel) with the result obtained using our improved method (right panel) at the same time. The original method produces a thicker dust layer than the new method, as expected for an overestimation of the drag force. With the improved method, the dust grains undergo more settling before the onset of the instability, but apart from the thickness of the dust layer, the other features of the instability remain.

\section{Conclusions}
\label{conclusions}

We have extended the semi-implicit time-integration method of \cite{LB14} for two-fluid dust/gas mixtures to account for net differences between the non-drag accelerations of the gas and the dust.  The improved method obtains the correct limiting velocity difference between the dust and the gas in the presence of differential accelerations even for time-steps that are much longer than the dust stopping time (i.e.\ $\delta t/t_{\rm s} \rightarrow \infty$). Due to the application of pair-wise forces, exact linear and angular momentum conservation are guaranteed. 

We have successfully applied the method to an accelerated \textsc{dustybox} test, demonstrating the accuracy of the method. We have also investigated the effect of the incorrect dust settling velocities produced by the earlier method in the generation of toroidal vortices in protoplanetary discs. Due to overestimation of the drag force, the earlier method produces slower settling of the dust particles, which gives rise to a somewhat thicker convective dust layer than that obtained with the new method. However, the onset of the instability and the character of the toroidal vortices are qualitatively unchanged.

\section*{Acknowledgments}
Figure \ref{disc} was created using SPLASH \citep{Pri07}, a SPH visualization tool publicly available at http://users.monash.edu.au/$\sim$dprice/splash.  

This work was supported by the STFC consolidated grant ST/J001627/1, and by the European Research Council under the European Community's Seventh Framework Programme (FP7/2007-2013 grant agreement no. 339248). This work used the DiRAC Complexity system, operated by the University of Leicester IT Services, which forms part of the STFC DiRAC HPC Facility (www.dirac.ac.uk). This equipment is funded by BIS National E-Infrastructure capital grant ST/K000373/1 and  STFC DiRAC Operations grant ST/K0003259/1. DiRAC is part of the National E-Infrastructure. This work also used the University of Exeter Supercomputer, a DiRAC Facility jointly funded by STFC, the Large Facilities Capital Fund of BIS and the University of Exeter.

\label{lastpage}
\end{document}